\documentclass{iau}
\usepackage{graphicx}

\title[Magnetic Field Structure of 37 Comae] %% give here short title %%
{Magnetic Field Structure and Activity of the He-burning Giant 37 Comae}

\author[Tsvetkova et al.]   %% give here short author list %%
{S. Tsvetkova$^1$, P. Petit$^2$, R. Konstantinova-Antova$^{1,2}$, M. Auri\`ere$^2$, G.A. Wade$^3$, C. Charbonnel$^{4,2}$, N.A. Drake$^{5,6}$}
\affiliation{$^1$Institute of Astronomy and NAO, Bulgarian Academy of Sciences, Bulgaria\\ email: {\tt stsvetkova@astro.bas.bg} \\[\affilskip]
$^2$IRAP, UMR 5277, CNRS and Universit\'e de Toulouse, France \\[\affilskip]
$^3$Department of Physics, Royal Military College of Canada, Ontario, Canada \\[\affilskip]
$^4$Geneva Observatory, University of Geneva, Switzerland \\[\affilskip]
$^5$Sobolev Astronomical Institute, St.~Petersburg State University, Russia \\[\affilskip]
$^6$Observat\'orio Nacional/MCTI, Rio de Janeiro, Brazil
}

\pubyear{2013}
\volume{302}  %% insert here IAU Symposium No.
\pagerange{xxx}
\jname{Magnetic Fields Throughout Stellar Evolution}
\editors{A.C. Editor, B.D. Editor \& C.E. Editor, eds.}
\begin{document}

\maketitle

\begin{abstract}

We present the first magnetic map of the late-type giant 37 Com. The Least Squares Deconvolution (LSD) method and Zeeman Doppler Imaging (ZDI) inversion technique were applied. The chromospheric activity indicators H$\alpha$, S-index, Ca\,{\sc ii} IRT and the radial velocity were also measured. The evolutionary status of the star has been studied on the basis of state-of-the-art stellar evolutionary models and chemical abundance analysis. 37 Com appears to be in the core Helium-burning phase.

\end{abstract}

\firstsection % if your document starts with a section,
\section{Introduction}

37 Com is the primary star of a wide triple system (Tokovinin 2008), but the synchronisation effect plays no role for its fast rotation and activity. Its significant photometric and Ca\,{\sc ii} H\&K emission variabilities were presented by Strassmeier et al. (1997; 1999) and de Medeiros et al. (1999) and interpreted as signatures of magnetic activity.

Observational data for 37 Com were obtained with two twin fiber-fed echelle spectropolarimeters -- Narval (2m TBL at Pic du Midi Observatory, France) and ESPaDOnS (3.6m CFHT). We have collected 11 Stokes V spectra for 37 Com in the period January 2010 -- July 2010. The Least Squares Deconvolution (LSD) multi-line technique was applied and the surface-averaged longitudinal magnetic field B$_l$ was computed using the first-order moment method (Donati el al. 1997; Wade et al. 2000). The Zeeman Doppler Imaging (ZDI) tomographic technique was employed for mapping the large-scale magnetic field of the star (Donati et al. 2006).

\section{Results}

%                                                 Figure
%-----------------------------------------------------------
  \begin{figure}[!htc]
    \centering
    \includegraphics[width=4cm, height=6cm]{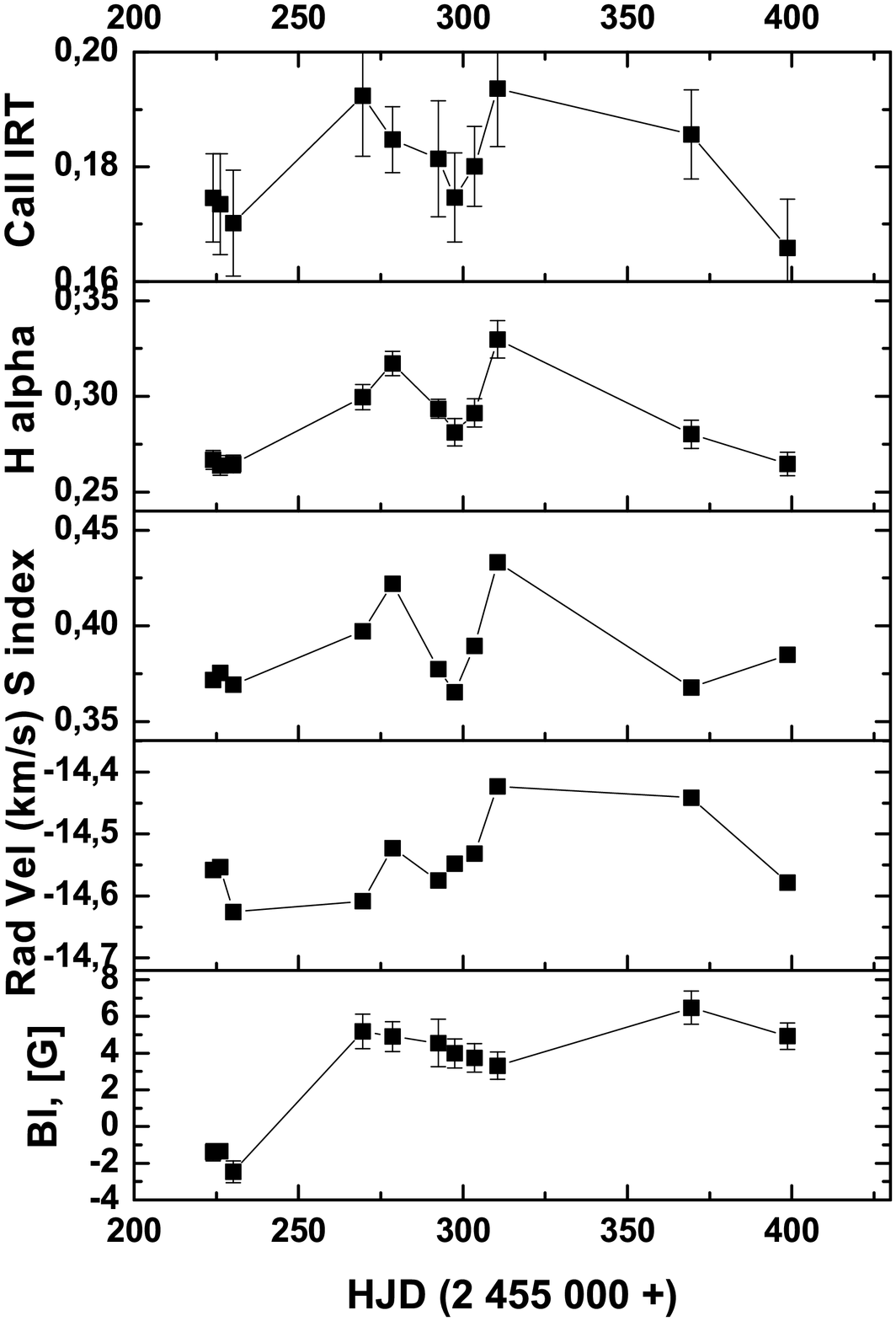}
    \includegraphics[width=3cm, height=6cm]{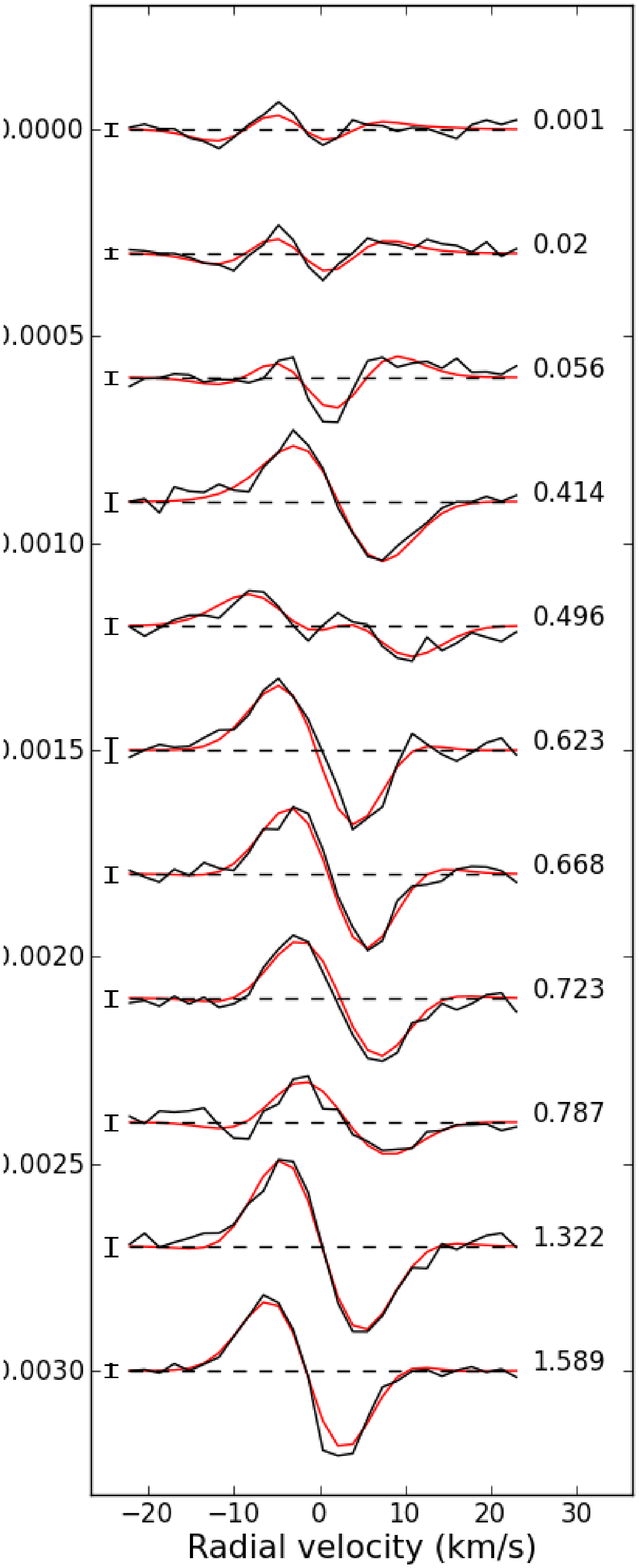}
    \includegraphics[width=6cm, height=6cm]{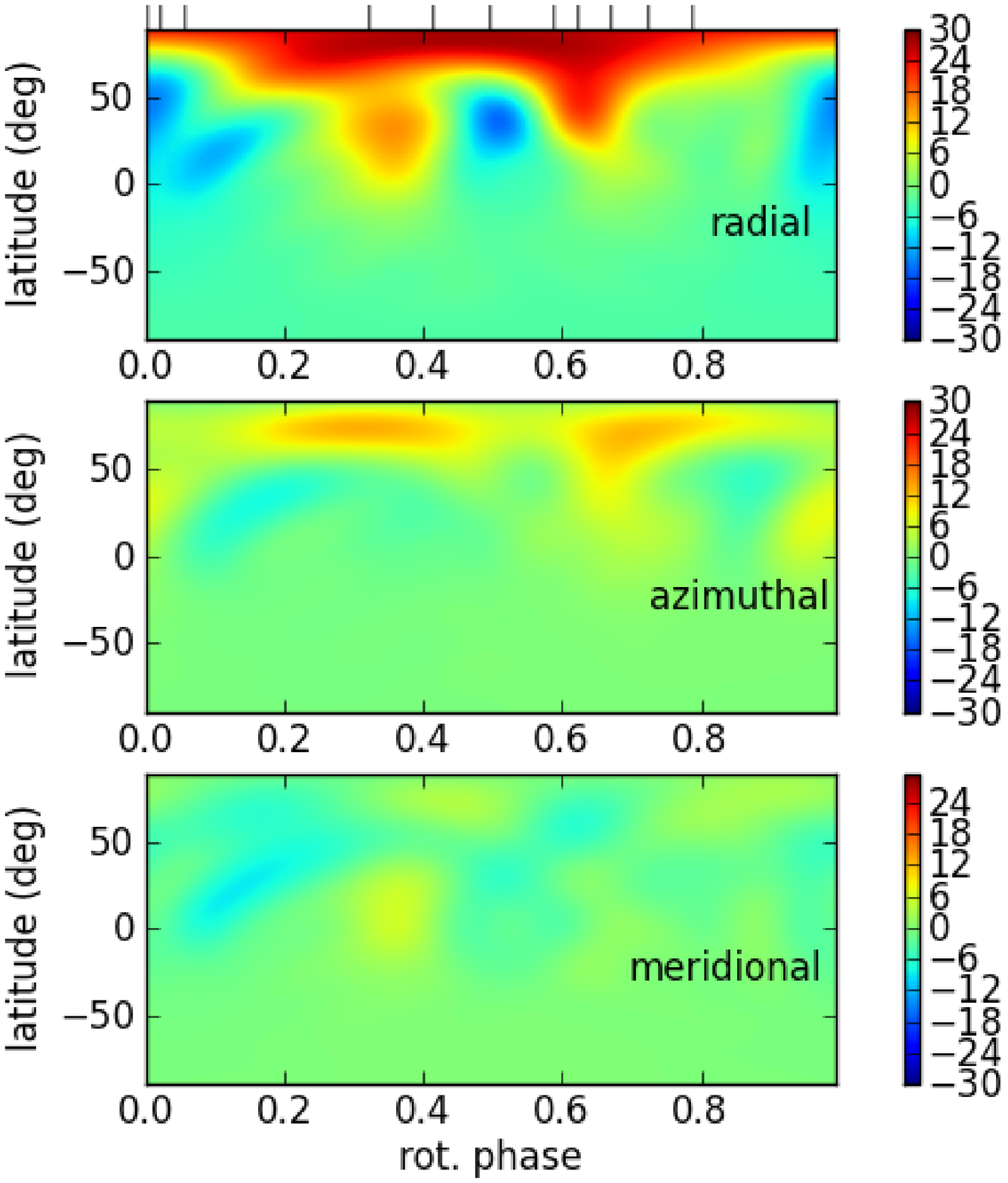}
     \caption{\textbf{Left:} From bottom to top -- simultaneous time variaions of B$_l$ with radial velocity (RV), S-index, H$\alpha$ and Ca\,{\sc ii} IRT (854.2 nm). \textbf{Center:} Normalized Stokes V profiles -- observed profiles (black); synthetic fit (red); zero level (dashes lines). The error bars are on the left of each profile. \textbf{Right:} The magnetic map of 37 Com. The magnetic field strength is in gauss. The vertical ticks on top of the radial map show the phases when there are observations.}
   \label{fig:ZDI}
   \end{figure}
%______________________________________________________________

There are significant variations of B$_l$ in the interval from -2.5 G to 6.5 G with at least one sign reversal during the observational period (Fig.~\ref{fig:ZDI} left). Also, radial velocity, S-index and line activity indicators H$\alpha$ and Ca\,{\sc ii} IRT (854.2 nm) show significant variations, and clear correlations with each other as well as the longitudinal field.

The ZDI mapping (Fig.~\ref{fig:ZDI} center and right) reveals that the large-scale magnetic field has a dominant poloidal component, which contains about 88\% of the reconstructed magnetic energy. The star has a differential rotation with the following parameters: $\Omega_{eq} = 0.06$ rad/d (the rotation rate at the equator) and $\Delta \Omega = 0.01$ rad/d (the difference in the rotation rate between the polar region and the equator) (Petit et al. 2002).

37 Com shows simpler surface magnetic structure than the fast rotators V390 Aur (Konstantinova-Antova et al. 2012) and HD 232862 (Auri\`ere et al. in prep.) and shows more complex structure than the slow rotators EK Eri (Auri\`ere et al. 2011) and $\beta$ Ceti (Tsvetkova et al. 2013), which are suspected of being descendants of Ap-stars.

The location of 37 Com on the Hertzsprung-Russell diagram was determined on the basis of state-of-the-art stellar evolution models (Charbonnel \& Lagarde 2010) and the mass  is found to be 5.25~$M_{\odot}$, in a good agreement with the literature. Synthetic spectra in the region containing $^{12}$CN and $^{13}$CN molecular lines were calculated and compared to our spectra in order to infer the $^{12}$C/$^{13}$C ratio. The best fit was achieved for $^{12}$C/$^{13}$C $ = 4.0$. From these results, it appears that 37 Com is in the core Helium-burning phase.

\begin{acknowledgements}
 STs thanks the contract BG05PO001-3.3.06-0047. G.A.W. is supported by an NSERC grant. N.A.D. thanks support of PCI/MCTI, Brazil, under the Project 302350/2013-6.
\end{acknowledgements}

\end{document}